\documentclass{jaa}
\usepackage{txfonts}
\usepackage{graphicx}
\newcommand{\newblock}{}
\usepackage[authoryear]{natbib}

\begin{document}

\title[]
{N(HI) and jet power/emission in AGNs }
\author[Wu et al. ]
       {Zhongzu Wu$^1$
       \thanks{e-mail:zzwu08@gmail.com}     Minfeng Gu$^2$  \& Ming Zhu$^3$\\
     $^1$   College of Science, Guizhou University,
                             Guiyang 550025, Guizhou, China. \\
                       $^2$ Shanghai Astronomical Observatory Shanghai, 200030, China\\
                        $^3$ National Astronomical Observatory Beijing, 100020, China
   }\maketitle
\label{firstpage}

\begin{abstract}
Neutral hydrogen (HI) 21 cm absorption has been detected against more and more powerful radio jets. In this work, based on the Guppta et al. 2006a sample, we present our preliminary study of the correlations between the HI column density N(HI) and the jet power, N(HI) versus the low frequency luminosity at 408MHz, and N(HI) versus the radio luminosity at 1400MHz.
\end{abstract}


\section{Introduction}

Relativistic plasma jets represent one of the possible ways in which the enormous energy released by active nucleus and it can also interact with and affect the interstellar medium \citep{morganti10}. It was suggested that the radio activity might be associated with presence of cold gas. The rotation axis of the cold gas is usually aligned
with the radio axis \citep{vangork89}. \cite{gupta06a} showed that the HI absorption for compact radio sources could be due to gas clouds accelerated by the radio jets, and it might be an evidence of jet-cloud interaction.

It is widely believed that the energy output from AGN is triggered by the supply of gas to the central engine, the dependence of N(HI) on radio luminosity could provide clues towards understanding this phenomenon \citep{gupta06a}. \cite{pihlst03} found an anti-correlation between N(HI) and  linear size of the jet and \cite{gupta06b} also found that N(HI) is related to the core prominence parameter and some other parameters.  In this work, we will try to study the correlations between the intrinsic low frequency radio luminosity and jet power with the HI column density around the jets ($H_{0}=70 \rm
{~km ~s^ {-1}~Mpc^{-1}}$, $\rm \Omega_{M}=0.3$, and $\rm
\Omega_{\Lambda} = 0.7$ were adopted and spectral
index $\alpha$ is defined as $f_{\nu} \propto \nu^{-\alpha}$).
\section{Sample selection and data collection}
\cite{gupta06a} obtained a sample of 96 radio sources from literatures
which is a heterogeneous superset of sources with available HI absorption observations.   The sample includes 27 gigahertz peaked spectrum (GPS), 35 compact steep spectrum (CSS), 13 compact flat
spectrum (CFS) and 21 large (LRG) sources, spaning a wide range of $\sim$ 6 orders of magnitude in luminosity at 5 GHz and z $\leq$ 1.4. We have collected the radio data for this sample at 1.4 GHz, 408MHz, 365 MHz, 333 MHz, 151 MHz, 178 MHz. The radio data at 1.4GHz were from VLA first survey or NVSS survey, the radio data at 408MHz, 151 MHz and other low frequency data were obtained from NASA/IPAC Extragalactic Database (NED).

Jet power is a fundamental parameter reflecting the energy transport to large spatial scales from the central engine by the radio jet.  In this work,
We estimated the jet power using the relation in
\cite{pun05}:
\begin{equation}
\rm{Q_{jet}=5.7\times10^{44}(1+z)^{1+\alpha}Z^{2}F_{151}} ~~
\rm{erg~s^{-1}}
\end{equation}
\begin{equation}
\rm
Z\approx3.31-3.65\times{[(1+z)^{4}-0.203(1+z)^{3}
+0.749(1+z)^{2}+
0.444(1+z)+0.205]^{-0.125}}
\end{equation}
where $F_{151}$ is the optically thin flux density from the lobes
measured at 151 MHz in units of Jy, and
$\alpha\approx1$ is assumed (see Punsly (2005) for more details).

\section{Correlation analysis}
 We investigated the correlations between N(HI) and the jet power, luminosity at 1.4 GHz and 408 MHz. The jet power $Q_{\rm jet}$ was calculated either directly from the radio flux at 151 MHz or from the extrapolated 151 MHz flux density from the available flux density at 160 MHz, 178 MHz, 330 MHz, 365 MHz or 408 MHz  assuming $\alpha\approx1$. The whole data table will be presented in a forthcoming paper \citep{wu11}. Due to the fact that the N(HI) are only given as upper limits in a large fraction of source due to non-detections, we used the astronomy Survival Analysis (ASURV) package (Isobe \& Feigelson 1990) in our correlation analysis when the upper limits were involved.

\begin{figure}
\centering
\includegraphics[width=15cm, angle=0]{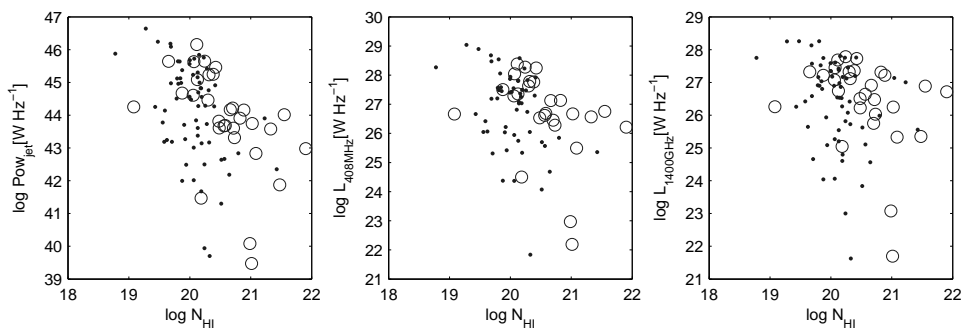}

\caption{left: The jet power versus N(HI); middle: low frequency radio luminosity at 408 MHz versus N(HI); right: the low frequency radio luminosity at 1400 MHz versus N(HI). Open circles represent the sources with detected N(HI) while the black dots are those sources only with upper limit of N(HI). }

\label{fig:pjet}
\end{figure}

 We find a significant negative correlation between $Q_{\rm jet}$ and N(HI) with a Spearman correlation coefficient of r=-0.707 at $\gg$ 99.99$\%  $ level for the sample with N(HI) detections (31 sources) and r=0.242 at  $\sim$ 98.23$\% $ level for the whole sample (95 sources) by using ASRUV package ( see Fig. \ref{fig:pjet}).
For the radio luminosity at 1400 MHz, we find a strong negative correlation (see Fig \ref{fig:pjet}) for detections (29 sources) with r=-0.5 at $\sim$ 99.2$\%$ level and no correlation for the whole sample (91 sources).  In case of the radio luminosity at 408 MHz, we find a significant negative correlation, r=-0,626 at $\gg$ 99.99$\%  $ level for detections and r=-0.199 at $\sim$ 92.4$\%  $ level for the whole sample (81 sources) ( see Fig  \ref{fig:pjet}) .
   The inclusion of the non-detections seems to significantly weaken the correlation, which was found to also affect the correlation between N(HI) and jet size in \cite{curran10}.

As the jet power and low frequency radio luminosity are normally believed
to indicate the intrinsic jet power or emission, our results may imply that the HI gas around powerful radio jet might be tightly related with radio jets. However this needs further investigations due to the small sample size, and only upper limits of N(HI) in many sources.

\noindent{{\textbf{Acknowledgements:}}} \\
\noindent{This work is supported by the NSFC grants (No. 10978009, 10703009, 10821302, 1083002, 11073039, and 10803015), and by the 973 Program (No. 2009CB824800). This research has made use of NASA/IPAC Extragalactic Database (NED),
which is operated by the Jet Propulsion Laboratory, California Institute
of Technology, under contract with National Aeronautics and Space
Administration.}

\label{lastpage}


\begin{thebibliography}{100}
\setlength{\itemsep}{0pt}
\bibitem[Curran \& Whiting (2010)]{curran10}
Curran, S. J. and Whiting, M. T. 2010, \newblock {\it Astrophys. J.}, {\bf 712}, 303
\bibitem[Gupta et al.(2006a)]{gupta06a}
Gupta, Neeraj., Salter, C. J., Saikia, D. J., Ghosh, T., and Jeyakumar, S., 2006a, \newblock {\it Mon. not. R. Astron. Soc.}, {\bf 373}, 972
\bibitem[Gupta et al.(2006b)]{gupta06b}
Gupta, Neeraj., and Saikia, D. J., 2006b, \newblock {\it Mon. not. R. Astron. Soc.}, {\bf 370}, 738
\bibitem[Isobe \& Feigelson (1990)]{Isobe90}
Isobe, T. and Feigelson, E. D. 1990, \newblock {\it Bull. Am. Astron. Soc}, {\bf 22}, 917
\bibitem[Morganti et al.(2010)]{morganti10}
Morganti, R., Holt, J., Tadhunter, C., and Oosterloo, T., 2010, IAU Symposium 267
\bibitem[Pihlstr$\ddot{o}$m et al.(2003)]{pihlst03}
Pihlstr$\ddot{o}$m, Y. M., Conway, J. E., and Vermeulen, R. C. 2003, \newblock {\it Astron. Astrophys.}, {\bf 404}, 871
\bibitem[Punsly (2005)]{pun05}
Punsly, B. 2005, \newblock {\it Astrophys. J.}, {\bf 623}, L9
\bibitem[Van Gorkom et al.(1989)]{vangork89}
van Gorkom, J. H., Knapp, G. R., Ekers, R. D., Ekers, D. D., Laing, R. A., and Polk, K. S. 1989, \newblock{\it Astron. J.}, {\bf 97}, 708
\bibitem[Wu et al.(2011)]{wu11}
Wu, Zhongzu, et al. 2011, \newblock{\it in preparation}


\end{thebibliography}
\end{document}